# Testing for Markovian character of transfer of fluctuations in solar wind turbulence on kinetic scales

Dariusz Wójcik[*] and Wiesław M. Macek[†]

*Institute of Physical Sciences, Faculty of Mathematics and Natural Sciences,
Cardinal Stefan Wyszyński University, Wóycickiego 1/3, 01-938 Warsaw, Poland
and Space Research Centre, Polish Academy of Sciences, Bartycka 18A, 00-716 Warsaw, Poland*

We apply statistical analysis to search for processes responsible for turbulence in physical systems. In our previous studies, we have shown that solar wind turbulence in the inertial range of large magnetohydrodynamic scales exhibits Markov properties. We have recently extended this approach on much smaller kinetic scales. Here we are testing for the Markovian character of stochastic processes in a kinetic regime based on magnetic field and velocity fluctuations in the solar wind, measured onboard the Magnetospheric Multiscale (MMS) mission: behind the bow shock, inside the magnetosheath, and near the magnetopause. We have verified that the Chapman-Kolmogorov necessary conditions for Markov processes is satisfied for *local* transfer of energy between the magnetic and velocity fields also on kinetic scales. We have confirmed that for magnetic fluctuations, the first Kramers-Moyal coefficient is linear, while the second term is quadratic, corresponding to drift and diffusion processes in the resulting Fokker-Planck equation. It means that magnetic self-similar turbulence is described by generalized Ornstein-Uhlenbeck processes. We show that for the magnetic case, the Fokker-Planck equation leads to the probability density functions of the kappa distributions, which exhibit global universal *scale invariance* with a linear scaling and lack of intermittency. On the contrary, for velocity fluctuations, higher order Kramers-Moyal coefficients should be taken into account and hence scale invariance is not observed. However, the nonextensity parameter in Tsallis entropy provides a robust measure of the departure of the system from equilibrium. The obtained results are important for a better understanding of the physical mechanism governing turbulent systems in space and laboratory.

## I. INTRODUCTION

Turbulence consists of phenomena that, notwithstanding progress in numerical simulations, remain a challenge for natural sciences, even for simple fluids [1]. This is all the more so for magnetized plasma with magnetohydrodynamic (including Hall) simulations, but physical mechanisms responsible for irregular behavior are still not transparent [2]. Fortunately, collisionless solar wind plasma can be considered a natural laboratory for investigating these complex dynamical systems [3]. In particular, fluctuations of magnetic fields play an important role in space plasmas [e.g., 4,5]. The classical spectrum of Kolmogorov [6] follows a power law with slope exponent $-5/3$ for isotropic incompressible turbulence in ordinary fluids and Kraichnan [7] type spectrum with an even smaller slope of $-3/2$ in magnetized media.

One should underline that in a Markov process, given an initial probability distribution function (PDF), the transition to the next stage can be fully determined. It is also interesting that one can prove and demonstrate the existence of such Markov processes based on experimental data [8]. Namely, without relying on any assumptions or models for the underlying stochastic process, we are able to extract the equation of Markov process directly from the measurements of times series. Hence this Markov approach appears to be a bridge between the statistical and dynamical analysis of complex physical systems. There is a substantial evidence based on statistical analysis that turbulence exhibits Markov properties [8–13]. We have already proved that magnetic and velocity fluctuations have Markovian features in the inertial range of hydromagnetic scales [10,11]. In this case, the characteristic spectrum appears to be roughly close to the standard Kolmogorov [6] power-law type with exponent $-5/3 \approx -1.67$; see Fig. 1 of Ref. [10].

On the other hand, the presence of the classical $-5/3$ or $-3/2$ at small scales seems to be rather exceptional. In fact, based on the highest millisecond time resolution data available in the Magnetospheric Multiscale (MMS) mission, we have noted clear breakpoints in the magnetic energy spectra in various regions of the Earth's magnetosheath: (a) behind the bow shock (BS), (b) inside the magnetosheath (SH), and (c) near the magnetopause (MP) before leaving the magnetosheath. Namely, we have observed that the magnetic spectrum steepens at some critical points in the kinetic regime of scales: from $-5/2$ above the ion gyrofrequency till $-7/2$ or even $-11/2$ (or $-16/3$) above the Taylor-shifted frequency (related to the electron skin depth of above 20–25 Hz), but for velocity spectrum similar to Kolmogorov [6]; see Ref. [14]. Some higher slopes for the magnetic spectrum have also been observed for an electron gyroscale [15] and for whistler turbulence [e.g., 16,17].

---

[*]Contact author: dwojcik@cbk.waw.pl
[†]Contact author: macek@uksw.edu.pl, macek@cbk.waw.pl; http://www.cbk.waw.pl/~macek

Therefore, it is certainly interesting to investigate the Markov property of turbulence outside the inertial region of magnetized plasma on small scales, when the slopes are consistent with kinetic theory [e.g., 18]. It should also be noted that based on the measurements of magnetic field fluctuations in the Earth's magnetosheath gathered onboard the MMS mission, we have recently extended this statistical analysis to much smaller scales, where kinetic theory should be applied [19,20]. Therefore, here we compare the characteristics of both magnetic field and velocity fluctuations behind the bow shock, inside the magnetosheath, and near the magnetopause. In this paper, we present the results of our comparative analysis for the following cases of transfer of fluctuations: velocity-to-velocity, velocity-to-magnetic, and magnetic-to-velocity, confirming the local character of the transfer of cascading eddies. We also check whether the Fokker-Planck (FP) equation is suitable for the processes responsible for solar wind turbulence and whether their solutions agree with experimental PDFs in selected regions of the magnetosheath.

Section II provides a brief description of the MMS mission and the analyzed data. Section III outlines mathematical stochastic and statistical methods for processes in scale (Sec. III A), including the necessary Chapman-Kolmogorov (CK) condition in Sec. III B. This enables estimating the Kramers-Moyal (KM) coefficients and checking the validity of Pawula's theorem leading to the FP equation in Sec. III C, given the stationarity (Appendix A) of the detrended time series (Appendix B). The results of our analysis are presented in Sec. IV, which demonstrates that at least for magnetic fluctuations, the solutions of the FP equation are in good agreement with empirical PDFs also on small kinetic scales, with a universal global *scale invariance*. Finally, Sec. V emphasizes the significance of stochastic processes in relation to turbulence in space plasmas, which exhibit Markovian features across the kinetic domain.

## II. DATA

The MMS mission was launched in 2015 to investigate plasma processes in the magnetosphere and the solar wind plasma especially on small scales [21]. Our recent analysis has encompassed increments in each vector component of the magnetic field, denoted as $\mathbf{B} = (B_x, B_y, B_z)$, within the Geocentric Solar Ecliptic (GSE) coordinate system, as displayed together with the corresponding spectra in Fig. 1 of Ref. [19]. This vector field $\mathbf{B}$ have been derived from measurements collected by the MMS 1 spacecraft, positioned just beyond the bow shock region of the Earth. We have successfully demonstrated the presence of Markovian characteristics, where the magnetic turbulence occurs, even at significantly reduced kinetic scales. Interestingly, our observations also reveal a remarkable uniformity in these Markovian features across all components. We have also explored analogous characteristics of the magnitude of this vector field, $B = |\mathbf{B}|$, in three different regions within the magnetosphere: (a) behind the bow shock, (b) within the magnetosheath, and (c) in proximity to the magnetopause, as depicted in Fig. 1 of Ref. [20].

Consequently, we now aim to analyze the magnetic field strength $B = |\mathbf{B}|$, as well as the ion plasma velocity $V = |\mathbf{V}|$ magnitudes inside the magnetosheath and in its close vicinity. These regions as well as the spacecraft trajectory have been depicted in Fig. 1 of Ref. [14]. To explore the potential global scale invariance of the fluctuations, we have opted for the same three time interval samples as those presented in Table 1 of Ref. [14]. For a magnetic field, in cases (a) near the bow shock and (c) near the magnetopause, with time spans of approximately 5 min and 1.8 min each, we have used BURST-type observations obtained from the FluxGate Magnetometer (FGM) sensor. This sensor has the highest resolution of $\Delta t_B = 7.8$ ms (roughly 128 samples per second), providing data sets with 37 856 and 13 959 data points, respectively. Conversely, in case (b) between the bow shock and the magnetopause, we have the available FAST-survey type data at substantially lower resolution of $\Delta t_B = 62.5$ ms (16 samples per second). This data set consists of a much longer interval, spanning 3.5 h, and contains a total of 198 717 data points.

In addition, in the analysis of ion plasma velocity magnitude $V = |\mathbf{V}|$, we use data obtained through measurements from the Dual Ions Spectrometer (DIS) instrument. These measurements have a lower time resolution; specifically within the BURST-type observations, the time resolution $\Delta t_V$ for ion measurements is set at 150 ms (about 6.5 samples per second), and 30 ms (33 samples per second) for electrons. In the FAST-survey mode, the instrument provides data snapshots at regular intervals of $\Delta t_V = 4.5$ s. All the data sets considered in this study are available through Ref. [22], while the complete description of the MMS spacecraft instruments is specified in Ref. [21].

## III. METHODS OF DATA ANALYSIS

As usual, the examination of statistical properties of a turbulent system is performed across various scales. Kolmogorov [6] has already postulated that the isotropic turbulence within the inertial range should be linked to the flow velocity increment of $\tau$. We also suppose that fluctuations occurring at a larger scale $\tau$ shift to smaller and smaller scales, until the dissipation scale is reached [19,20]. Therefore, here we employ the increments (fluctuations) of a given parameter, either the characteristic magnetic field denoted by $X := B = |\mathbf{B}|$ or the ion velocity denoted by $X := V = |\mathbf{V}|$. Naturally, such increments are typical scale-dependent complexity measures, which characterize the behavior of a turbulent system at a given time $t$ and at each timescale (lag) $\tau$, as given by

$$x_\tau(t) := \delta x(t, \tau) = X(t + \tau) - X(t), \quad (1)$$

i.e., the difference in magnetic field or flow velocity between points separated by a time interval $\tau$.

In consequence, stochastic fluctuations can be perceived as a stochastic process, governed by the $N$-point joint (transition) conditional probability density function (denoted as cPDF), $P(x_1, \tau_1 | x_2, \tau_2; \ldots; x_N, \tau_N)$, where $P(x_i, \tau_i | x_j, \tau_j) =$ $\frac{P(x_i, \tau_i; x_j, \tau_j)}{P(x_j, \tau_j)}$ is a conditional PDF, with the joint PDF $P(x_i, \tau_i; x_j, \tau_j)$ and a marginal PDF $P(x_j, \tau_j)$. By employing these joint PDFs, the correlations between scales can also be determined, which illustrates how the complexity is linked across different scales. Nevertheless, the reasoning can easily be inverted by applying the *Bayes theorem* by the equation, for $P(x_j, \tau_j) \neq 0$,

$$P(x_i, \tau_i | x_j, \tau_j) = \frac{P(x_j, \tau_j | x_i, \tau_i) P(x_i, \tau_i)}{P(x_j, \tau_j)}. \quad (2)$$

When $\tau$ is scaled by a factor $\lambda \in \mathbb{R}^+$, then it is said that the increments are globally scale invariant, if scaling occurs with the unique scaling exponent $\beta$, i.e., $x_{\lambda\tau}(t) = \lambda^\beta x_\tau(t)$, with $\beta$ independent of scale $\tau$.

### A. Process in scale

The power spectral densities (PSDs) (as shown in the lower parts of Figs. 2–4 in Ref. [14]) serve as a statistic to scrutinize the scale-dependent behavior of turbulent fluctuations and are analogous to the examination of the autocorrelation function. In this study, we explore a somewhat expanded application of stochastic processes by means of the Markovian approach. Essentially, the stochastic process is said to be *Markovian* if, for $0 < \tau_1 < \tau_2 < \cdots < \tau_N$, it holds that

$$P(x_1, \tau_1 | x_2, \tau_2) = P(x_1, \tau_1 | x_2, \tau_2; \ldots; x_N, \tau_N). \quad (3)$$

In particular, if the data satisfy the Markov condition from larger to smaller scales, this condition is also satisfied in the reverse direction from smaller to larger scales, as stated previously by the Bayes theorem in Eq. (2); see Ref. [8].

### B. Chapman-Kolmogorov condition and locality

The generalization of Eq. (3) is called the *Chapman-Kolmogorov* (CK) *equation* (condition),

$$P(x_1, \tau_1 | x_2, \tau_2) = \int_{-\infty}^{+\infty} P(x_1, \tau_1 | x', \tau') P(x', \tau' | x_2, \tau_2) dx', \quad (4)$$

where $(\tau_1, \tau', \tau_2)$ is a set of timescale parameters, such as $\tau_1 < \tau' < \tau_2$. We have checked that it is satisfied for the Markov turbulence on inertial scales [10]. Importantly, this serves as the very essential requirement for a stochastic process to exhibit Markovian properties.

If the condition of Eq. (4) is fulfilled, then the transition PDF from the scale $\tau_2$ to $\tau_1$ can be decomposed into two sequential transitions: first from $\tau_2$ to $\tau'$, and next from $\tau'$ to $\tau_1$. Hence, in the context of a turbulent cascade, fulfillment of this condition for all considered $(\tau_1, \tau', \tau_2)$ implies the existence of a *local* transfer mechanism in the cascade. When considering fluctuations in time, one can imply that the transfer process is local in scale. Nonetheless, if the root mean square (RMS) (in a discrete case, $x_{\rm RMS} = [\frac{1}{n} \sum_i x_i^2]^{1/2}$) of ion velocity increments is considerably smaller than the mean velocity $\langle v_{\rm sw} \rangle$ of the solar wind flow, under Taylor's hypothesis [23], then temporal variations at a fixed position are understood as spatial variations. In this scenario, the local transfer in scale can be understood as being directly linked to the local transfer in wave-vector space.

We also aim to examine the interactions between ion velocity and magnetic field modes, which provide input into the transfer of energy between these two quantities. These interactions should be interpreted as a statistical dependence between $b_\tau(t, \tau)$ and $v_\tau(t, \tau)$. To incorporate this transfer of fluctuations between different quantities, we introduce the generalized CK condition as

$$\begin{aligned} P(x_1, \tau_1 | y_2, \tau_2) &= \int_{-\infty}^{+\infty} P(x_1, \tau_1 | y', \tau') P(y', \tau' | y_2, \tau_2) dy' \\ &\sim \int_{-\infty}^{+\infty} P(x_1, \tau_1 | x', \tau') P(x', \tau' | y_2, \tau_2) dx', \end{aligned} \quad (5)$$

where the intermediate-scale quantity varies, but yields very similar results, allowing for the use of only one version of Eq. (5). This also enables us to investigate whether the transfer of fluctuations between two quantities exhibits a local or nonlocal character. Specifically, if the empirical cPDFs align with those calculated from Eq. (5), then the transfer of fluctuations can be broken down into smaller steps (with the intermediate-scale $\tau'$), implying that the transfer in the cascade has a local character.

Consequently, the differential form of the CK equation (4) is called the *Kramers-Moyal* (KM) expansion and is given by

$$-\frac{\partial P(x_\tau, \tau | x'_\tau, \tau')}{\partial \tau} = \sum_{k=1}^{\infty} \left(-\frac{\partial}{\partial x_\tau}\right)^k \\ \times [D^{(k)}(x_\tau, \tau) P(x_\tau, \tau | x'_\tau, \tau')], \quad (6)$$

where the coefficients $D^{(k)}(x_\tau, \tau)$, called Kramers-Moyal (KM) coefficients, are given by

$$D^{(k)}(x_\tau, \tau) = \frac{1}{k!} \lim_{\tau \to \tau'} \frac{1}{\tau - \tau'} M^{(k)}(x_\tau, \tau, \tau'), \quad (7)$$

with

$$M^{(k)}(x_\tau, \tau, \tau') = \int_{-\infty}^{+\infty} (x'_\tau - x_\tau)^k P(x'_\tau, \tau' | x_\tau, \tau) dx'_\tau, \quad (8)$$

which can be obtained by extrapolation (piecewise linear regression model). Equations (7) and (8) show that the drift and diffusion coefficients can be expressed in the form of the first and second moments of the cPDFs $P(x'_\tau, \tau' | x_\tau, \tau)$ in the small time interval limit. In this way, one can find the KM coefficients for the increments $x_\tau$.

In general, the KM expansion (6) involves infinitely many evolution terms. Often the first- and second-order KM coefficients are different from zero, and hence statistically significant, while the third-, fourth-, and higher-order coefficients usually exhibit a tendency to gradually approach zero. The important *Pawula's* theorem [24] states that if the fourth-order KM coefficient is equal to zero, then $D^{(k)}(x_\tau, \tau) = 0$ for $k \geqslant 3$, and the series is limited to the second order.

### C. Fokker-Planck and Langevin equations

In this case, the differential KM expansion (6) is presented in a reduced form called the *Fokker-Planck* (FP) equation,

determining the evolution of the transition probability [24]:

$$-\frac{\partial P(x_\tau, \tau | x'_\tau, \tau')}{\partial \tau} = \left[ -\frac{\partial}{\partial x_\tau} D^{(1)}(x_\tau, \tau) + \frac{\partial^2}{\partial x_\tau^2} D^{(2)}(x_\tau, \tau) \right] \times P(x_\tau, \tau | x'_\tau, \tau'), \quad (9)$$

where the first and second terms are responsible for the *drift* and *diffusion* processes, respectively. One should note the negative sign which arises from the direction of time evolution, moving from larger to smaller scales, as we have previously assumed.

Consequently, a complementary approach with the following *Langevin* equation (using the *Itô's* definition) emerges:

$$-\frac{\partial x_\tau}{\partial \tau} = D^{(1)}(x_\tau, \tau) + \sqrt{D^{(2)}(x_\tau, \tau)} \, \Gamma(\tau), \quad (10)$$

for the $\delta$-correlated Gaussian white noise $\Gamma(\tau)$, which satisfies the following normalization conditions: mean $\langle \Gamma(\tau) \rangle = 0$ and a correlation $\langle \Gamma(\tau) \Gamma(\tau') \rangle = 2\delta(\tau - \tau')$, where $\delta$ is a Dirac delta function [13]. The stochastic nature of the fluctuations across different scales is encapsulated by the diffusion coefficient within the Langevin framework.

Equivalently, Langevin equation (10) can be rewritten in a more natural form of stochastic differential equation (SDE),

$$dx_\tau(\tau) = h(x_\tau, \tau) d\tau + g(x_\tau, \tau) dW(\tau), \quad (11)$$

where, again, $h(x_\tau, \tau) = D^{(1)}(x_\tau, \tau)$ is a drift term and $g(x_\tau, \tau) = \sqrt{D^{(2)}(x_\tau, \tau)}$ modulates a diffusion term. Here, $\{W(\tau) | \tau \geqslant 0\}$ is a scale-related Wiener process (Brownian motion). Generally, we say that $\{W(\tau)\}$ is a Brownian motion in scale if the following conditions are satisfied: (i) $W(\tau = 0) = 0$, (ii) mean $<W(\tau)> = 0$ and variance $\text{Var}[W(\tau)] = \sigma^2 \tau$, and (iii) $\{W(\tau)\}$ has stationary and independent increments. Note that the Markov property is implied by the presence of independent increments. There is an equivalence between the FP and Langevin dynamics, in a way that the PDF of a stochastic process whose dynamics is governed by the Langevin equation satisfies the FP given by Eq. (9); see the proof in Ref. [25].

Notably, any process $x_\tau(\tau)$ generated with Eq. (11) is a continuous diffusion process. Such a diffusion process refers to a continuous-time stochastic process with (almost surely) continuous sample paths having the Markov property. Actually, a fundamental example of a continuous diffusion process is a Wiener process. Moreover, the process $x_\tau(\tau)$ generated by Eq. (11) also satisfies the Lipschitz condition, i.e., for any function $f : \mathbb{R} \to \mathbb{R}$, and $x_1, x_2 \in \mathbb{R}$, there exists a constant $\epsilon > 0$ such that $|f(x_1) - f(x_2)| \leqslant \epsilon |x_1 - x_2|$.

On the other hand, in terms of the probabilities $\mathbb{P} : \mathcal{F} \to \mathbb{R}^+$, where $\mathcal{F}$ is an event space, we can say that the continuous process $x_\tau(\tau)$, generated by the Langevin equation, fulfills a continuity Lindenberg's (Dynkin's) condition, if, for any $\epsilon > 0$,

$$\lim_{\tau \to 0^+} \frac{\mathbb{P}[|x_\tau(\tau)| > \epsilon | X(\tau) = x]}{\tau} = 0. \quad (12)$$

Deviation from this continuity criterion implies a lack of smoothness of the process, signifying the presence of discontinuous events such as jumps. Consequently, for the process with nonvanishing higher-order KM coefficients, the Lindenberg's continuity condition of Eq. (12) can be rewritten as a continuity condition in terms of conditional moments $M^{(k)}$, for $k \geqslant 0$ and any $\epsilon > 0$,

$$\lim_{\tau \to 0^+} \frac{\mathbb{P}[|x_\tau(\tau)| > \epsilon | X(\tau) = x]}{\tau} \leqslant \frac{M^{(k)}(x_\tau, \tau)}{\epsilon^k \tau}, \quad (13)$$

which can be proven using the Chebyshev inequality; see [26].

The Langevin equation (11) [characterized by vanishing higher-order ($k \geqslant 3$) KM coefficients] generates continuous sample paths. However, complex systems often manifest nonstationary dynamics, leading to discontinuous sample paths in the corresponding time series [25]. This poses a challenge when employing the Langevin approach. Apparently, distinct features such as heavy tails or some abrupt large jumps may imply the existence of discontinuous jump components [27]. Models with jumps have been employed to capture this randomness [28]. Their primary challenge yet involves estimating parameters defining jumps and their distribution sizes, along with addressing path discontinuities in processes sampled at discrete intervals. Remarkably, this nonparametric estimation enables one to examine potential nonlinearities in drift, diffusion, and the intensity of the discontinuous *jump* component. This component can be related to the nonvanishing higher-order KM coefficients. Therefore, a SDE that describes a stochastic *jump-diffusion* process is of the following form [see [27]]:

$$dx_\tau(\tau) = h(x_\tau, \tau) d\tau + g(x_\tau, \tau) dW(\tau) + \xi(x_\tau, \tau) dJ(\tau), \quad (14)$$

where $J(\tau)$ represents a timescale-homogeneous Poisson jump process, with a jump rate $\lambda(x_\tau)$, a size $\xi(x_\tau, \tau) \sim \mathcal{N}(0, \sigma_\xi^2)$, and a jump amplitude $\sigma_\xi^2$. In this process, the diffusion coefficient and a jump characteristics contribute to the second-order KM coefficient. Notably, all unknown functions and coefficients can be directly derived from the empirical data. This approach is suitable for both stationary and nonstationary time series, where discontinuous jump components are present, which would need further investigation; cf. Ref. [27].

## IV. RESULTS

We focus on Markovian characteristics at much finer millisecond scales. Certainly, this allows one to extend our analysis beyond the inertial range [10,11], with a particular emphasis on magnetic field fluctuations, as has already been analyzed in Refs. [19,20].

In our endeavor to better understand the turbulence mechanisms within space plasma, we analyze fluctuations not only of the magnetic field $B = |\mathbf{B}|$, but also of ion velocity $V = |\mathbf{V}|$. Namely, we have now applied this approach to the small-scale, in cases (a) and (c), and the medium-scale, in case (b), fluctuations of $B$, while for $V$ fluctuations, to medium scales in cases (a) and (c), and a higher scale in case (b).

### A. Stationarity

It is necessary to validate the stationarity of the data time series under investigation, as described in Sec. III A. To evaluate this feature, we have used the statistical tests, namely, the *Augmented Dickey-Fuller* (ADF) test [29], as well as the *Kwiatkowski-Phillips-Schmidt-Shin* (KPSS) test [30], as described in Appendix A.

Using the ADF test, we have determined that in cases (a) and (b), for both variables $B = |\mathbf{B}|$ and $V = |\mathbf{V}|$, respectively, the corresponding $p$ values are less than 0.01. This signifies the rejection of the null hypothesis. Consequently, the time series of magnetic field strength and ion velocity can be considered stationary. Nevertheless, for case (c), where the dataset is substantially smaller for both variables, the computed $p$ values equal 0.154 and 0.3705, for $B$ and $V$, respectively, indicating the inability to reject the null hypothesis. These values point towards the presence of some nonstationary component in the time series, implying a time-dependent structure with possibly fluctuating variance over time. Using the differencing method (lag-1 difference), we could get rid of this nonstationarity. Then the ADF test results in a $p$ value of $<0.01$, which leads again to the rejection of the null hypothesis.

Next, we have also employed the KPSS test on both variables $B = |\mathbf{B}|$ and $V = |\mathbf{V}|$. It turned out that in all of the analyzed cases (a)–(c), the resulting $p$ values are $<0.01$, which means that we reject the null hypothesis. This implies that our time series does have a unit root [which was already confirmed by the ADF test in case (c)], and it is rather nonstationary. In such a case, when the ADF test concludes stationarity, while the KPSS test suggests nonstationarity (presence of a unit root), this means that the series are difference stationary. Hence, to achieve stationarity, a straightforward lag-1 differentiation is usually sufficient. Indeed, after applying this operation, the resulting $p$ values are greater than 0.1 in all of the cases, demonstrating that we do not have sufficient evidence to reject the null hypothesis. Consequently, this signifies that both time series of magnetic field strength $B = |\mathbf{B}|$ and ion velocity $V = |\mathbf{V}|$ can be considered stationary.

### B. Chapman-Kolmogorov condition

The basis of adopting the Markov analysis relies on the assumption that the data follow Markovian characteristics. This can be subjected to direct verification through the definition given by Eq. (4). To establish the applicability of a Markov processes framework in our context, we can check the required CK condition, using a method described in Sec. III B. Although, to validate this necessary condition, we briefly discuss the integration limits of the right-hand side of this equation, which formally extends over the set of real numbers. We see that $(\tau_1, \tau', \tau_2)$ represents a set of fluctuation scales.

Note that $\tau_1$ has been chosen to be approximately three times greater than $\Delta t_B$, which determines a lower-bound scale in relation to the Einstein-Markov scale [31]. This lower bound represents a finite step size introducing a coarse-grained structure to the evolution of $\tau$ across scales, from the largest to the smallest scale. As a result, the Markov process can be interpreted as a stochastic model that effectively represents this continuous coarse-grained process.

Within Eq. (4), the integration is performed over $x'$ representing fluctuations at scale $\tau'$. In the calculations, it is possible to determine bounds for $x'$ based on various quantiles, such as the 5% and 95%, or more robustly, the 1% and 99% quantiles, or even more rigorously. These values are derived from time series at timescales $\tau'$. Formally, for a sufficiently large sample, the result of the integration should remain consistent, independently of the chosen limits. This implies that the chosen limits should be properly extensive. By using such a parametric case study, a consistent plateau can be identified for the well-defined limits of the integration.

Hence, using a methodology described in Sec. III B, we have obtained the empirical cPDFs from the data, denoted as $P_E(x_1, \tau_1 | x_2, \tau_2)$, as presented in Figs. 1–3, plotted as continuous contours with the matching various colors (a) near the bow shock, (b) inside the magnetosheath, and (c) near the magnetopause. They are compared there with the cPDFs, which are solutions of the CK equation (4), labeled as $P_{CK}(x_1, \tau_1 | x_2, \tau_2)$, and shown as the black dashed contours. Such a comparison is done under the specific parameter set $(\tau_1, \tau', \tau_2)$, such as $\tau' = \tau_1 + \Delta t_X$, $\tau_2 = \tau_1 + 2\Delta t_X$, for $\tau_1 < \tau' < \tau_2$. Here, $\Delta t_X$ again denotes a sampling time encompassing each variable $B$ and $V$, for cases (a)–(c), which are comprehensively described below. Typically, the contour graphs provide a two-dimensional (2D) representation of 3D data.

The results of this comparative analysis for magnetic field increments $x_i = b_i$ $\forall_i$ performed at various kinetic scales are displayed in Fig. 2 of Ref. [20], as compared with Fig. 1 of Ref. [11] for the region of large magnetohydrodynamic (MHD) scales. The analogous comparative analysis is now seen in Fig. 1 using ion velocity increments $x_i = v_i$ at various scales, compared with Fig. 2 of Ref. [11]. More precisely, in cases (a) and (c), with lower time separation $\Delta t_V = 0.15$ s, we have taken into account the following scales: $\tau_1 = 0.3$ s, $\tau' = \tau_1 + \Delta t_V = 0.45$ s, and $\tau_2 = \tau_1 + 2\Delta t_V = 0.6$ s. In case (b), with much lower resolution $\Delta t_V = 4.5$ s, we have used $\tau_1 = 9$ s, $\tau' = \tau_1 + \Delta t_V = 13.5$ s, and $\tau_2 = \tau_1 + 2\Delta t_V = 18$ s. Again, the isolines shown in the graphs indicate decreasing levels of cPDFs for velocity increments $v_i$ from the central region of each plot. They are as follows: in case (a), 0.4, 0.16, 0.09, 0.04, 0.02, 0.01, and 0.006; in case (b), 0.2, 0.031, 0.023, 0.015, 0.009, 0.005, 0.003; while in case (c), 0.4, 0.145, 0.09, 0.07, 0.04, 0.03, and 0.015. We observe that in the central parts, the contour lines corresponding to $P_E(\cdot | \cdot)$ and $P_{CK}(\cdot | \cdot)$ overlap quite well in cases (a) and (b), and somewhat less in case (c). We see pronounced irregularities caused by the limited data available for estimating PDFs near the edges of the plots, especially in case (c). Hence, in such regions, the explicit verification of the suitability of the CK equation (4) might not be directly possible. Nevertheless, towards the central parts of the contours, we can observe that the CK equation (4) is sufficiently well fulfilled in the first two cases, and approximately satisfied in the third case.

In addition, to analyze the superposed *locality* of fluctuations, as thoroughly described in Sec. III B, we have derived the empirical cPDFs $P_E(x_1, \tau_1 | y_2, \tau_2)$ and $P_E(y_1, \tau_1 | x_2, \tau_2)$, as well as the solutions of a generalized CK equation (5), given by cPDFs $P_{CK}(x_1, \tau_1 | y_2, \tau_2)$ and $P_{CK}(y_1, \tau_1 | x_2, \tau_2)$, respectively. One should note that the time stamps of the resolution for the $B$ and $V$ variables for *in situ* measurements by the MMS should receive greater attention. We have noticed slight shifts in the time stamps, but they are very comparable (up to several milliseconds), and thus we have concluded that these shifts are negligible. To begin with, to achieve an "equilibrium" between the two analyzed variables, we have employed methods of upsampling (linear interpolation) and downsampling (decimation) of data, as elaborated in Appendix B.

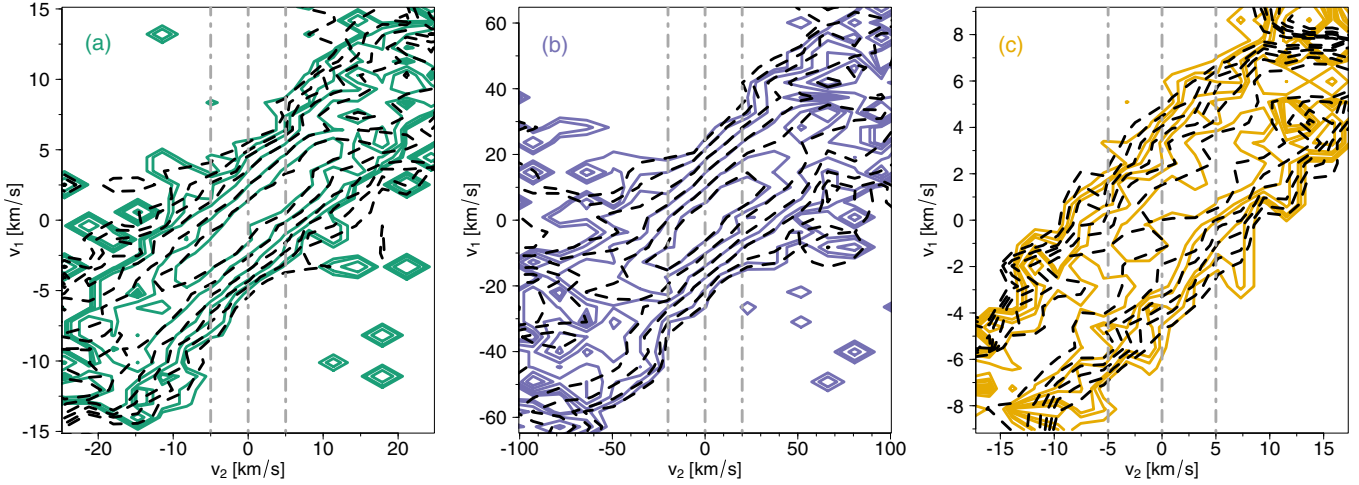

FIG. 1. A comparison between the experimental contours of cPDFs, $P_E(v_1, \tau_1|v_2, \tau_2)$ (depicted as solid various-colored curves), and the reconstructed CK contours of cPDFs, $P_{CK}(v_1, \tau_1|v_2, \tau_2)$ (shown as dashed black curves), for different timescales using the CK condition (4). Data from the MMS mission for ion velocity $V = |\mathbf{V}|$, in three cases: (a) near the bow shock, (b) inside the magnetosheath, and (c) near the magnetopause.

Namely, in case (a), with $\Delta t_B = 0.0078$ s and $\Delta t_V = 0.15$ s, we have performed a downsampling on the magnetic field variable to obtain a consistent time resolution of $\Delta t_{BV} \sim 0.15$ s, resulting in $\sim 2000$ data points. In turn, in case (b), with $\Delta t_B = 0.0625$ s and $\Delta t_V = 4.5$ s, because of the risk of substantial loss of information through a straightforward downsampling (loss of too many, e.g., 1 per 72 observations), we have opted for linear interpolation on the variable of ion velocity $V$ joined with downsampling on magnetic field variable $B$. This has yielded a time resolution of roughly $\Delta t_{BV} \sim 2.25$ s, with a dataset consisting of 5500 points. To enhance the clarity of the results in case (c), we have applied the same joint methodology to have a longer (initially severely constrained) variable of velocity $V$. This results in a satisfactory time resolution of $\Delta t_{BV} \sim 0.075$ s, with about 1500 data points, that facilitates a generation of the specific set of scale parameters $(\tau_1, \tau', \tau_2)$. Similar to the previous one-variable cases, $\tau_1 = 0.3$ s in case (a), $\tau_1 = 4.5$ s in case (b), and $\tau_1 = 0.15$ s in case (c), while $\tau' = \tau_1 + \Delta t_{BV}$ and $\tau_2 = \tau_1 + 2\Delta t_{BV}$.

Since the exchange between magnetic and kinetic energy is through magnetic field line stretching, expressed as $\mathbf{b} \cdot \nabla \mathbf{v} \cdot \mathbf{b}$, then physically the intermediate-scale quantities are $x' = b'$ and $y' = b'$ in Eq. (5) in the transfer velocity-to-magnetic field ($x_1 = b_1$ and $y_2 = v_2$) and magnetic-to-velocity field ($x_1 = v_1$ and $y_2 = b_2$), respectively. Although both $x'$ and $y'$ as intermediate-scale quantities are mathematically rigorous, only one selected version of Eq. (5) should be used, as mentioned in Sec. III B. The results for both cases of the comparison of the exchange of energy between the velocity and the magnetic fields on kinetic scales are presented as contours in Figs. 2 and 3, correspondingly.

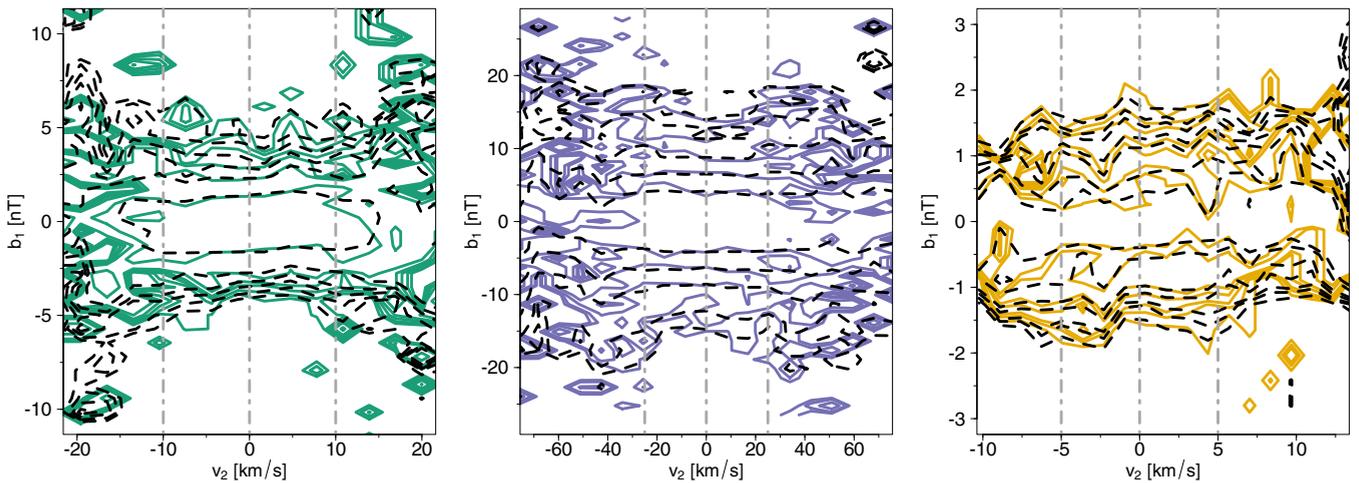

FIG. 2. A comparison between the experimental contours of cPDFs, $P_E(b_1, \tau_1|v_2, \tau_2)$ (depicted as solid various-colored curves), and the reconstructed CK contours of cPDFs, $P_{CK}(b_1, \tau_1|v_2, \tau_2)$ (shown as dashed black curves), for different timescales using the generalized CK condition (5) with $x' = b'$. Data from the MMS mission for a $V$ to $B$ transfer of fluctuations, in the same three analyzed cases.

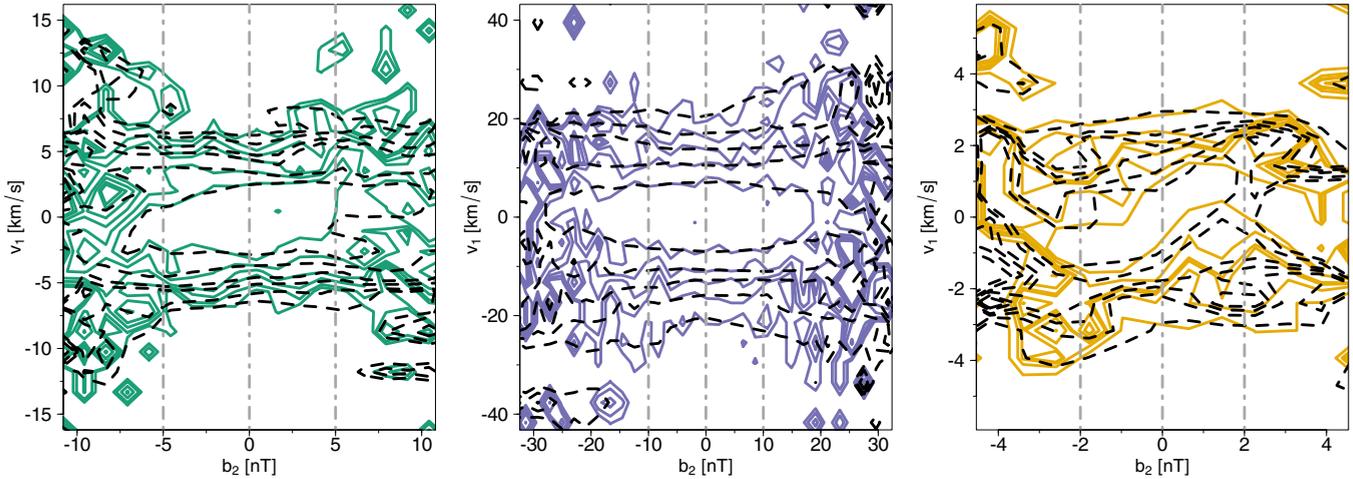

FIG. 3. A comparison between the experimental contours of cPDFs, $P_E(v_1, \tau_1 | b_2, \tau_2)$ (depicted as solid various-colored curves), and the reconstructed CK contours of cPDFs, $P_{CK}(v_1, \tau_1 | b_2, \tau_2)$ (shown as dashed black curves), for different timescales using the generalized CK condition (5) with $y' = b'$. Data from the MMS mission for a $B$ to $V$ transfer of fluctuations, in the same three analyzed cases.

Further, the isolines displayed in the graphs represent decreasing levels of the cPDFs for magnetic field increments $b_i$, from the center of the plots. For the velocity-to-magnetic case, we show the isolines originating from the middle of the plots set as follows: 0.3, 0.14, 0.07, 0.045, 0.035, 0.025, 0.018 in case (a); 0.2, 0.055, 0.03, 0.017, 0.011, 0.007, 0.004 in case (b); and 0.4, 0.25, 0.16, 0.14, 0.11, 0.09, 0.06 in case (c). Since the energy released from the magnetic field should be equal to the energy obtained by the velocity field, the transfer of velocity-to-magnetic fluctuation should be approximately the same (with opposite signs) as the magnetic-to-velocity exchange, similarly to what has also been computed for MHD scales; see Figs. 5 and 6 in Ref. [11].

Figures 2 and 3 show that both experimental and theoretical contours align quite well in the central parts and to a slightly lesser extent in the outer regions of cases (a) and (b). However, these fits are less pronounced than in the transfer of fluctuations of the same quantity. Additionally, in case (c), one sees a discrepancy even in the center parts, especially in the case of magnetic-to-velocity transfer. We think that these irregularities are primarily related to the limited amount of the data available here. Anyway, these observations suggest that the fulfillment of the generalized CK equation (5) is rather tangible in cases (a) and (b), but admittedly may not be conclusive in case (c). Please note the gray vertical dashed lines in each contour plot (Figs. 1–3) across all investigated cases.

Further, to provide supplementary validation of the proposed CK condition (4) and the generalized CK condition (5), we have also examined the cross sections through the 3D histograms of the cPDF for specific increments $x_i = v_i$ or $x_i = b_i$, for the respective $v_i$–$v_i$, $b_i$–$v_i$, and $v_i$–$b_i$ transfers of fluctuations, as illustrated in Figs. 4–6, respectively. The approximated (discretized) fixed values of each parameter are given at the uppermost part of each plot. As can be seen, the slices (cross sections) through the empirical cPDFs, $P_E(\cdot | \cdot)$, plotted as colored filled circles, align quite well with the slices through the theoretical curves, $P_{CK}(\cdot | \cdot)$, shown with black dashed curves, yielding a very good agreement with the CK condition of Eq. (5) in all of the examined cases; cf. [20], Fig. 3 (for the transfer of magnetic-to-magnetic fluctuations, $x_i = b_i$ and $x'_i = b_i$). The slight deviations in tails observed on a few graphs can be attributed to the limited data used for the unsupervised binning method (see Appendix B). For the fixed values of $b_2 < 0$ (nT) and $b_2 > 0$ (nT), we see (on top of each plot) a slight shift and truncation of the PDFs, which is prominent especially in cases (a) and (b), and to a lesser degree, in case (c).

It is important to highlight that our extensive comparisons across various parameter sets $(\tau_1, \tau', \tau_2)$ have yielded convincing results. In particular, we have found that the CK condition (4) holds true for $x_\tau = b_\tau$ as $\tau$ increases [20], Fig. 3. When $\Delta t_B$ attains somewhat higher values, indicating that $\tau_2$ becomes sufficiently large, the CK condition remains satisfied, but sometimes both cPDFs no longer exhibit dependence on $b_2$. Nevertheless, this broader insight in analyzed space regions suggests that the turbulent cascade usually has Markov properties also on kinetic scales.

Admittedly, in Fig. 4, for the velocity-to-velocity transfer fluctuations, we see that the alignment is slightly less clear than that observed in the case of magnetic-to-magnetic fluctuations in Fig. 3 of Ref. [20]. The most accurate fits are seen in case (a), but minor deviations are encountered in the tails of each PDF. Rather good fits are observed in case (b), though numerical noise is present even in the central peaked part of each PDF. Admittedly, the least precise fitting emerges in case (c). In general, for fixed values of $v_2 < 0$ (km s$^{-1}$) and $v_2 > 0$ (km s$^{-1}$) (see top of each plot), the shapes of the PDFs can exhibit some asymmetry, which may be attributed to higher moments of PDF in Eq. (8). In the former case, the obtained PDF shape seems to be somewhat right-skewed, while in the latter situation, a tendency to a left-skewed shape is rather apparent. Additionally, the shapes in cases (a) and (b) exhibit greater peakedness, whereas in case (c), we observe somewhat smoother shapes. Apart from that, both of the first PDFs match quite well; also in case (c), the fits are fairly acceptable.

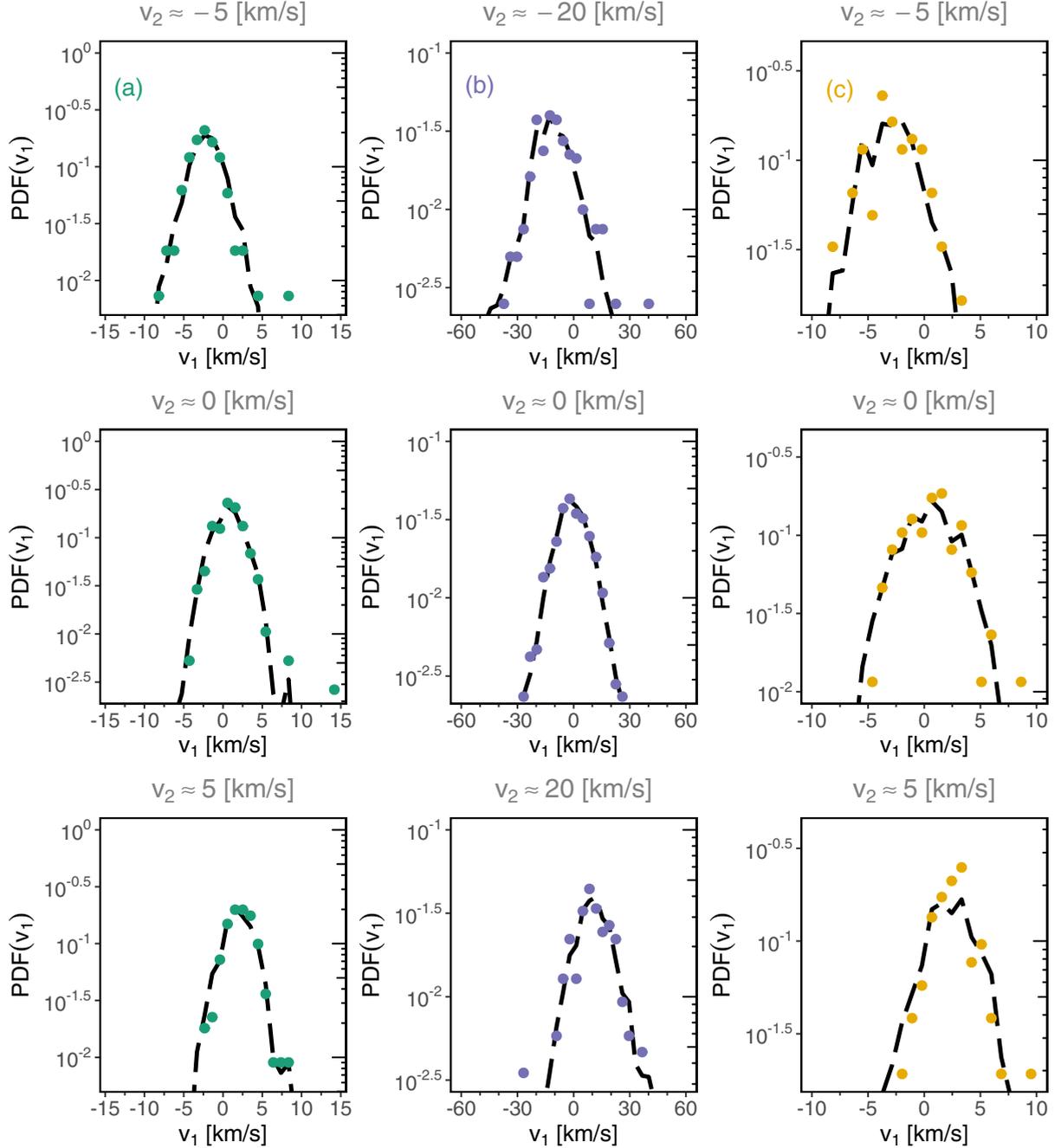

FIG. 4. Comparison of the experimental cross sections through the cPDFs $P_E(v_1, \tau_1|v_2, \tau_2)$ (colored filled circles) and $P_{CK}(v_1, \tau_1|v_2, \tau_2)$ (black dashed lines), under fixed ion velocity fluctuation $v_2$ (on top of each graph) for cases (a) behind the bow shock, (b) inside the magnetosheath, and (c) near the magnetopause. The parameter set $(\tau_1, \tau', \tau_2)$ is in agreement with that in Fig. 1.

It is also relevant to discuss the extent to which the CK condition given by Eq. (4) is satisfied for the $x_i = v_i$ variable. After a comprehensive analysis involving numerous parameter sets $(\tau_1, \tau', \tau_2)$ akin to the previous scenario, the following observations emerge. In case (a), this condition is approximately well satisfied up to $50\Delta t_V = 7.5$ s, corresponding to $\tau_2 = \tau_1 + 50\Delta t_V = 7.8$ s. Next, in case (b), it remains valid for scales up to $30\Delta t_V = 135$ s, which gives $\tau_2 = \tau_1 + 30\Delta t_V = 144$ s. In case (c), it is somewhat fulfilled, albeit up to $5\Delta t_V = 0.75$ s, namely, $\tau_2 = \tau_1 + 5\Delta t_V = 1.05$ s. Once again, our observations affirm the presence of Markovian properties within the turbulent cascade as the analysis is able to look at the domain of kinetic scales.

Afterwards, we consider the transfer of fluctuations between the two quantities: velocity-to-magnetic and magnetic-to-velocity. Here we present the cross sections on kinetic scales in Figs. 5 and 6, respectively, which should be compared with the corresponding Figs. 7 and 8 of Ref. [11] for MHD scales. Notably, the circles representing empirical cPDFs exhibit a reasonable fit with the slices through the solutions of the generalized CK of Eq. (5). The most accurate fit is seen in case (b), with a slightly less precise fit in case (a), while the most significant deviations are apparent in case (c). The deviations in the tails are present, which can be attributed to some outliers, which are more challenging to understand and potentially remove in this joint method for limited data. In addition, slight shifting and truncation of the PDFs are observed, in particular in case (c).

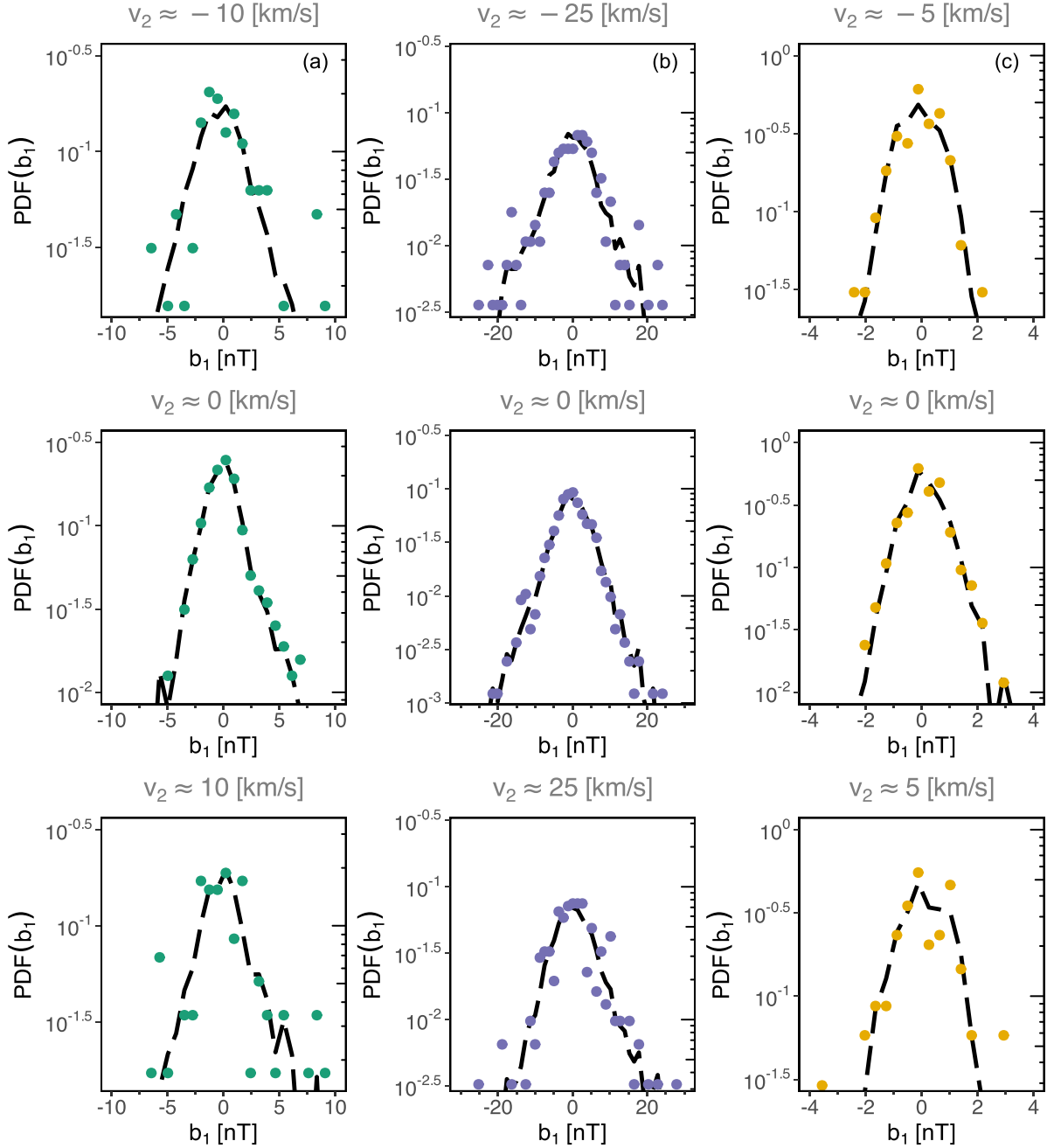

FIG. 5. Comparison of the experimental cross sections through the cPDFs $P_E(b_1, \tau_1 | v_2, \tau_2)$ (colored filled circles) and $P_{CK}(b_1, \tau_1 | v_2, \tau_2)$ (black dashed lines), under fixed ion velocity fluctuation $v_2$ (on top of each graph) for cases (a) behind the bow shock, (b) inside the magnetosheath, and (c) near the magnetopause. The parameter set $(\tau_1, \tau', \tau_2)$ is in agreement with that in Fig. 2.

This analysis suggests that the fulfillment of the appropriate CK condition (5) is at least approximately satisfied for the smallest considered range of scales available for testing. Our results provide supporting evidence that the Markov approach can be applied for the description of the turbulent cascade in solar wind turbulence. Therefore, the assumption of locality of the energy transfer in wave-vector space is valid, and thus

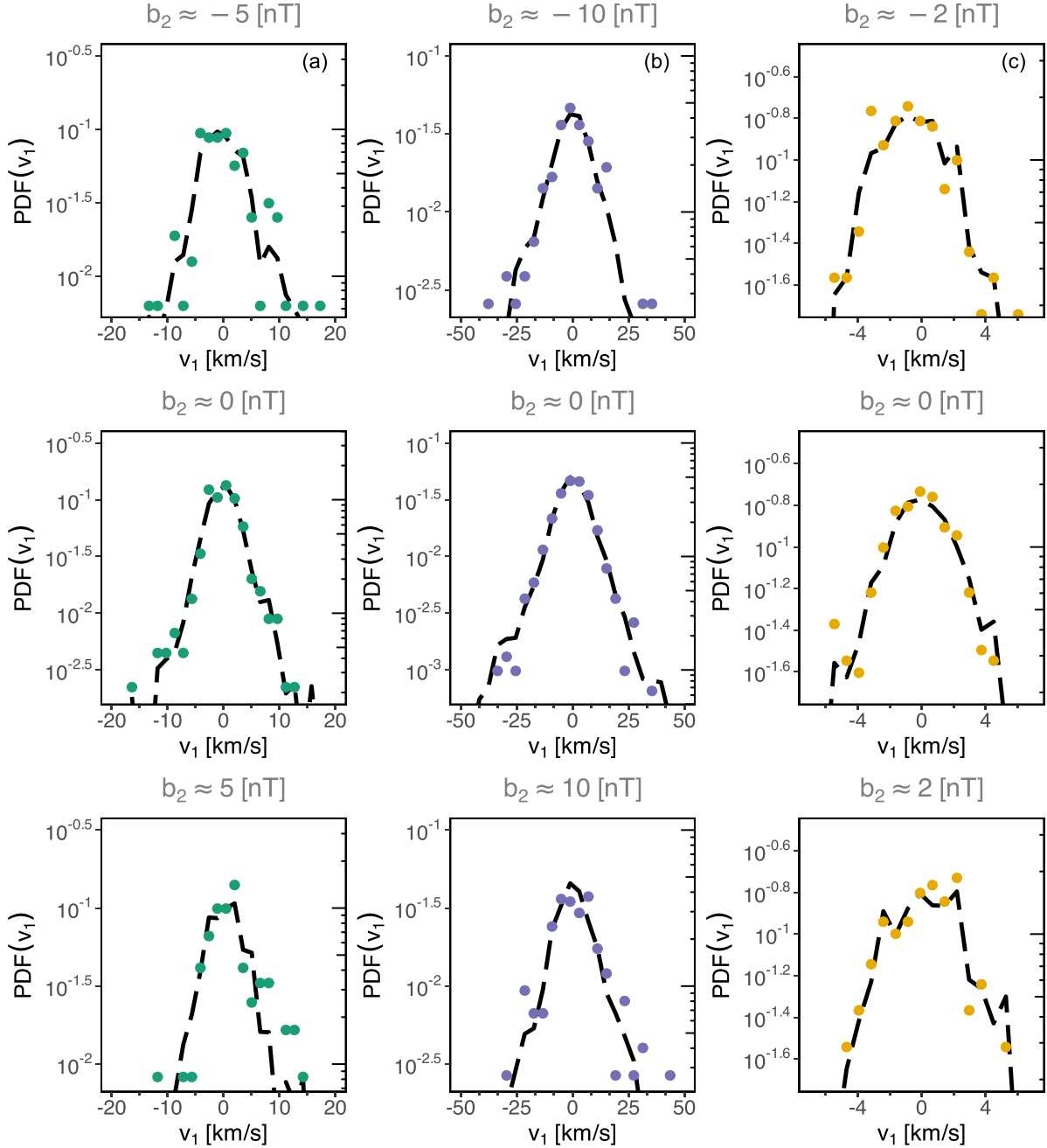

FIG. 6. Comparison of the experimental cross sections through the cPDFs $P_E(v_1, \tau_1|b_2, \tau_2)$ (colored filled circles) and $P_{CK}(v_1, \tau_1|b_2, \tau_2)$ (black dashed lines), under fixed magnetic field fluctuation $b_2$ (on top of each graph) for cases (a) behind the bow shock, (b) inside the magnetosheath, and (c) near the magnetopause. The parameter set $(\tau_1, \tau', \tau_2)$ is in agreement with that in Fig. 3.

our analysis extends the findings in the near-Earth space environment to small kinetic scales.

### C. Drift and diffusion coefficients

This further allows estimating the KM coefficients and checking the validity of Pawula's theorem as discussed in Sec. III C. Naturally, we have determined the KM coefficients of orders $k = 1, 2$, and 4, defined by Eq. (7). These coefficients bear crucial importance for the validation of Pawula's theorem, described in Sec. III B. The conventional approach employed for the determination of these coefficients involves the application of an extrapolation technique, for instance, the piecewise linear regression (mentioned in Sec. III B), to estimate the corresponding limits as $\tau \to \tau'$. It is important in this regard to recall the Einstein-Markov scale, below which the process is no longer Markovian. In our case, this derived scale is not too high, and the sampling rate is adequately high, thus not significantly affecting the determination of the KM coefficients. However, for $M^{(k)}(x_\tau, \tau, \tau')$ as a function of $\tau'$, we have noticed a consistent deviation from a linear fit for small values of $\tau'$ in Eq. (8), attributed to the Einstein-Markov scale (not shown here) that ought to be omitted when calculating the limits; see, e.g., [8], Fig. 14 and [12], Fig. 42. Therefore, we have directly computed these scaled central moments, using the obtained empirical cPDFs $P(x_1, \tau_1|x_2, \tau_2)$, for $0 < \tau_1 < \tau_2$. This systematic approach has yielded the empirical KM coefficients $D^{(k)}(x_\tau, \tau)$ obtained using point-by-point extrapolation of the $M^{(k)}(x_\tau, \tau, \tau')$ coefficients.

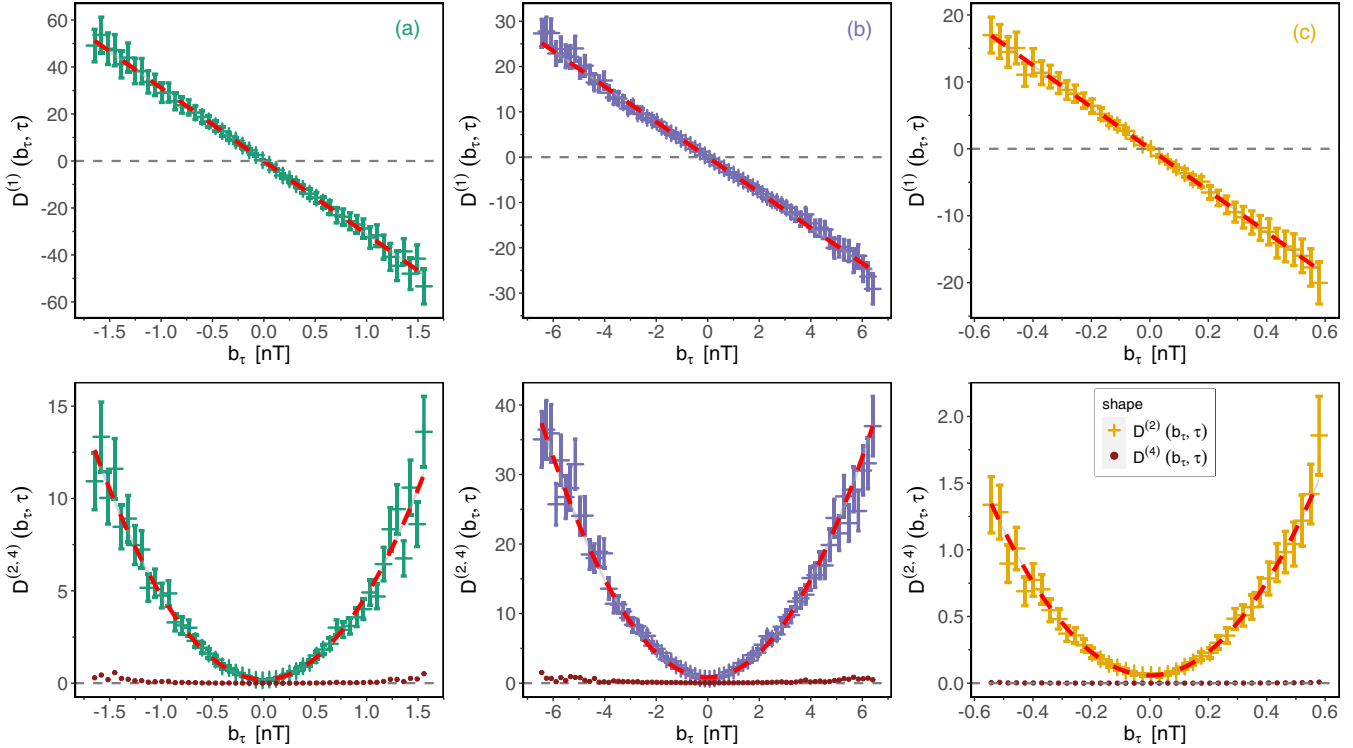

FIG. 7. The first and second Kramers-Moyal coefficients as functions of magnetic field increments $b_\tau$, for the magnetic field strength $B = |\mathbf{B}|$. Dashed red lines represent the optimal fits to the empirical values of $D^{(1)}(b_\tau, \tau)$ and $D^{(2)}(b_\tau, \tau)$, while $D^{(4)}(b_\tau, \tau) \approx 0$ (brown dots), maintaining the condition in agreement with Pawula's theorem, taken from Ref. [20].

In addition, since the errors of the number of occurrences $N(x'_\tau, x_\tau)$ of fluctuations $x'_\tau$ and $x_\tau$ can be resolved by $\sqrt{N(x'_\tau, x_\tau)}$, we have used an analogous reasoning to determine the errors of the scaled conditional moments $D^{(k)}(x_\tau, \tau)$, [see, e.g., [8]]. Consequently, Figs. 7 and 8 provide the first-order coefficient (in the upper segment), and the second- and fourth-order coefficients (in the lower segment) for $b_\tau$ and $v_\tau$ fluctuations, respectively. These graphs cover all three cases (a), (b), and (c), which are distinguished, as previously shown, by various colors. For each case, we also present the calculated confidence intervals.

The results depicted in Fig. 7 reveal a distinct pattern: the fit for the drift $D^{(1)}(b_\tau, \tau)$ adopts a linear relationship with respect to $b_\tau$, with negative slope, while diffusion $D^{(2)}(b_\tau, \tau)$ stands as a second-degree (parabolic) function of $b_\tau$. This behavior is consistent for $\Delta t_B = 0.0078$ s in cases (a) and (c), and $\Delta t_B = 0.0625$ s in case (b). Remarkably, our comprehensive analysis extends this fitting to significantly larger scales, up to $150\Delta t_B$, corresponding to $\tau_2 = \tau_1 + 150\Delta t_B$ for all three cases. This robust consistency suggests that fitting to lower-order polynomials for different $\Delta t_B$ is possible on kinetic scales.

We see that for increments of the magnetic field $B$, while the drift and diffusion coefficients remain nonvanishing, the fourth coefficient (brown dots) is approximately zero for all cases. Thus, the KM expansion of Eq. (6) truncates at $k = 2$ in cases (a)–(c), hence reducing to the FP equation. This results from Pawula's theorem, also ensuring the statistical continuity of the analyzed process. Therefore, the Markov process can effectively be described by this FP equation (9) (or, alternatively, by the Langevin equation (10); see, e.g., [32]).

In the corresponding Fig. 8, we can notice a pattern similar to that in Fig. 7, but with significantly less pronounced alignment. The fit for the drift $D^{(1)}(v_\tau, \tau)$ follows a clearly discernible linear relationship with respect to $v_\tau$, whereas the diffusion coefficient $D^{(2)}(v_\tau, \tau)$ exhibits a somewhat less apparent quadratic dependence on $v_\tau$. Some moderate deviations from observed behavior are seen, especially in cases (a) and (c), with $\Delta t_V = 0.15$ s, and to a lesser extent in case (c), with $\Delta t_V = 4.5$ s. Contrary to the magnetic case, we now see that the KM expansion of Eq. (6) does not stop for $k \geqslant 3$, suggesting that truncation of expansion at any finite lower order would result in some not entirely credible PDFs, which cannot be reconstructed using the FP or Langevin equations [24]. Basically, since the Pawula's theorem is not well satisfied, we can expect that to maintain the Markovian property of the process for velocity fluctuations, considering the higher-order coefficients in the KM equation (6) (possibly for $k \to \infty$) would be necessary. Our results indicate that the observed time series somewhat deviates from the class of continuous diffusion processes (discussed in Sec. III C), which may point to the influence of some jump events within the underlying stochastic process [27], which would need some more investigation.

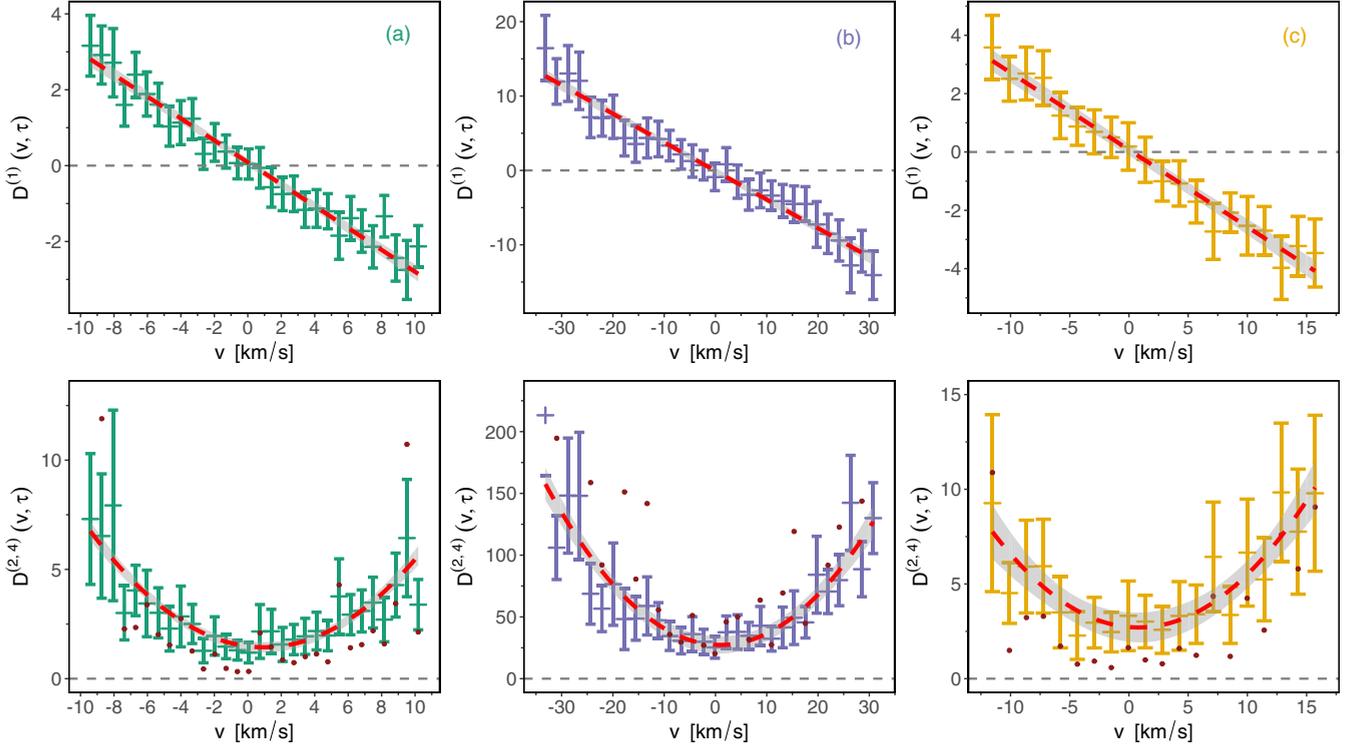

FIG. 8. The first and second Kramers-Moyal coefficients as functions of ion velocity increments $v_\tau$, for the ion velocity $V = |\mathbf{V}|$. Dashed red lines represent the optimal fits to the empirical values of $D^{(1)}(v_\tau, \tau)$ and $D^{(2)}(v_\tau, \tau)$, with $D^{(4)}(v_\tau, \tau) \neq 0$, i.e., not maintaining the Pawula's theorem.

### D. Distributions and entropies

It is worth noting that for the stationary solutions of the FP given by Eq. (9), the probability current must be constant. Hence, the following formula can be derived (see Sec. 5.2 of Ref. [24]):

$$p_s(x, \tau) = N_0 \exp\left[\int_{-\infty}^{x} \frac{D^{(1)}(x')}{D^{(2)}(x')} dx' - \ln D^{(2)}(x)\right]. \quad (15)$$

In this way, with a linear $D^{(1)}(b, \tau) = -a_1(\tau)b$ and a quadratic $D^{(2)}(b, \tau) = a_2(\tau) + b_2(\tau)b^2$, for the magnetic case, the stationary solution of Eq. (15) becomes a PDF of the *Kappa* distribution, given by [19]

$$p_s(b, \tau) = \frac{N_0}{\left[1 + \frac{1}{\kappa}\left(\frac{b}{b_0}\right)^2\right]^\kappa}, \quad (16)$$

where $\kappa = 1 + \frac{a_1(\tau)}{2b_2(\tau)}$ is a shape parameter and $b_0 = \sqrt{\frac{a_2(\tau)}{b_2(\tau)+a_1(\tau)/2}}$ is a scale parameter. The distribution function has to be normalized, $\int_{-\infty}^{\infty} p_s(b, \tau) db = 1$. We obtain

$$N_0 = \frac{1}{\mathcal{B}(\kappa - \frac{1}{2}, \kappa) b_0 \sqrt{\pi \kappa}}, \quad (17)$$

where $\mathcal{B}$ is a mathematical Beta function.

In particular, the Kappa distribution approaches the normal Gaussian distribution for large values of $\kappa$. This follows from the properties of the Beta function in the form $\mathcal{B}(\kappa_1, \kappa_2) = \frac{\Gamma(\kappa_1)\Gamma(\kappa_2)}{\Gamma(\kappa_1+\kappa_2)}$, and $\lim_{k\to\infty} \frac{\Gamma(k+\alpha)}{\Gamma(k)k^\alpha} = 1$, for $\alpha \in \mathbb{R}$. Fixing $\alpha = \frac{1}{2}$, and because $\Gamma(\kappa - \frac{1}{2})$ grows asymptotically at the same rate as $\frac{\Gamma(\kappa)}{\sqrt{\kappa}}$, the limit of Eq. (17) is simply given by $\lim_{\kappa\to\infty} N_0 = \frac{1}{b_0\sqrt{\pi}}$. Hence, as $\kappa \to \infty$ in Eq. (16), the well-known formula for the normal density distribution is obtained,

$$\lim_{\kappa\to\infty} p_s(b, \tau) = \frac{1}{b_0\sqrt{\pi}} e^{-\left(\frac{b}{b_0}\right)^2}, \quad (18)$$

with the mean value $\mu = 0$ and the standard deviation $\sigma = \frac{b_0}{\sqrt{2}}$, characterizing this symmetric PDF with extremely small tails. In addition, from the asymptotic expansion using the Stirling's factorial formula, one can show that in this case (with the respective third and fourth central moments $\mu_3$ and $\mu_4$), not only the third $\kappa_3$ (skewness) but also the fourth $\kappa_4 = \mu_4/\sigma^4 - 3$ (excess kurtosis) moments both approach zero.

On the other hand, nonzero $\kappa_3$ measures the possible asymmetry for left ($\kappa_3 < 0$) or right ($\kappa_3 > 0$) skewed PDF, but the value of $\kappa_4$ says how heavily the tails differ from a normal distribution. The Kappa distribution as a special case of the Pearson's type-*IV* family (i.e., a Pearson's type-VII distribution) have been used in numerous studies of solar wind and is important in space plasma physics [e.g., [33]]. This type of PDF is symmetric ($\kappa_3 = 0$) with heavy tail and peaked shape, measured by a positive $\kappa_4$ (leptokurtic) or negative $\kappa_4$ (platokurtic) kurtosis, which exhibits the deviation from the standard "bell" shape of the normal (mesokurtic) Gaussian distribution ($\kappa_4 \approx 0$).

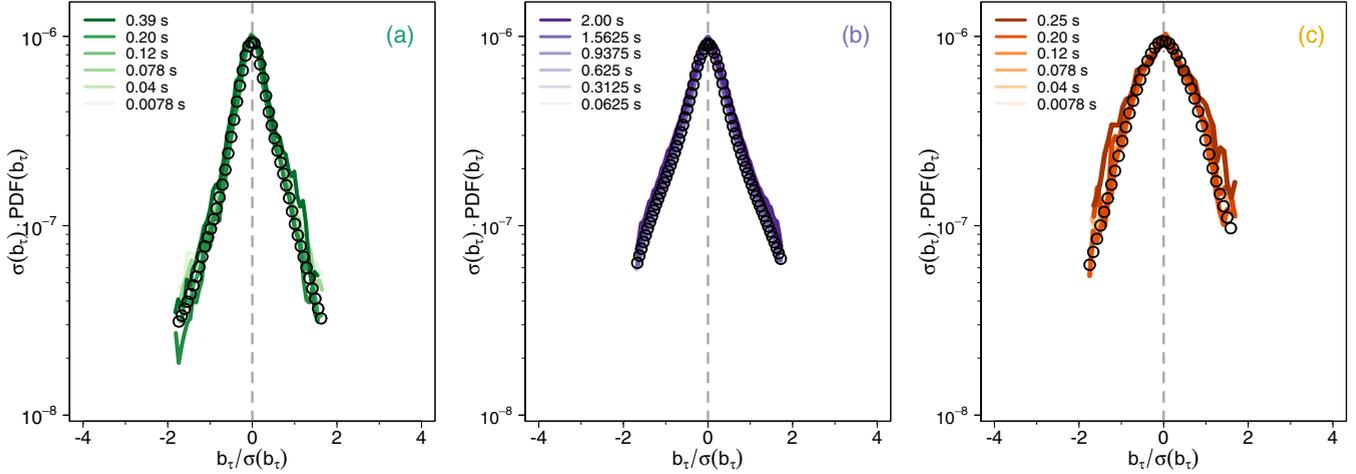

FIG. 9. Comparison between the rescaled empirical cPDFs (various continuous colored lines) and the stationary solutions of the FP equations (black open circles), considering the magnetic field magnitude $B = |\mathbf{B}|$ in all three cases. The global scale invariance can be observed.

Further, as is known, in classical statistical mechanics, Boltzmann additive entropy describes a system in equilibrium. On the other hand, the higher-order nonadditive nonextensive entropy is needed for a nonlinear nonequilibrium system, such as the solar wind at small scales [34]. Tsallis [35] introduced a generalization of the Boltzmann's entropy (with $k_B$ the Boltzmann constant),

$$S_q[p(x)] = \frac{k_B}{q-1}\left\{1 - \int_{-\infty}^{+\infty}[p(x')]^q dx'\right\}, \quad (19)$$

where $p(x)$ is a given probability distribution, and any real number $q$ is a nonextensive Tsallis parameter, which measures the degree of nonextensivity. For $q \to 1$, the classical Boltzmann-Gibbs information (additive) entropy is recovered, given by $-k_B \int p(x') \ln p(x') dx'$.

To unify power-law behaviors with a proper information measure, we use Tsallis entropy given by Eq. (19). Namely, to relate the derived Kappa distribution, with the Tsallis $q$-distribution function, a simple transformation can be used, $\kappa = 1/(1-q)$; see, e.g., Ref. [36]. The numerical results of fitting the MMS empirical magnetic data to the given distributions and determining the relevant parameters of Eq. (16) are as follows: $\kappa = 1.5179$, $x_0 = 1.9745$, and $N_0 = 0.68438$ for $B$ in case (a); $\kappa = 1.3758$, $x_0 = 2.6955$, and $N_0 = 0.34375$ in case (b); with $\kappa = 3.5215$, $x_0 = 1.7313$, and $N_0 = 1.1866$ in case (c).

Because the $\kappa$ parameter depends on $a_1(\tau)$ and $b_2(\tau)$, the KM coefficients can be related to the supposed nonextensive character of the fluctuations. The particular values of this $\kappa$ parameter can be attributed to the nonextensivity parameter $q$, which characterizes the generalized Tsallis entropy. In this situation, it is given by $q = \frac{a_1(\tau)}{a_1(\tau) + 2b_2(\tau)}$ and is equal to 0.341 in case (a), 0.273 in case (b), and 0.716 in case (c). The extracted values of the $\kappa$ and $q$ parameters provide robust measures of the departure of the system from equilibrium.

Based on magnetic MMS data, we have already obtained Kappa distributions for various scales, including the leptokurtic shapes for moderate scales (for the smallest scale values 7.8 ms described by a peaked shape close to the Dirac $\delta$ function) up to approximately the standard normal Gaussian distribution for the scale two orders of magnitude larger [20], Fig. 6. For relatively higher levels, the distributions still exhibit the peaked shape with high kurtosis, which refers to tails of the PDF. For velocity fluctuations, higher-order Kramers-Moyal coefficients should also be taken into account, with more complicated PDFs with possible slight asymmetry shown in Figs. 4–6.

Figure 9 illustrates the universal global scale invariance of the PDFs up to kinetic scales of (a) $\tau \sim 0.4$ s, (b) $\tau \sim 2$ s, and (c) $\tau \sim 0.25$ s, correspondingly [20]. This monoscaling scale invariance is obtained by rescaling cPDFs by their corresponding standard deviations $\sigma$ for each case, as explained in Refs. [19,20]. This demonstrates the exceptional agreement between empirical cPDFs and the stationary solutions (denoted by black open circles) of the FP equation (9) provided by the analytic formula of Eq. (16). One can only mention here that these results imply that the mean $n$-order structure function $S_n(\tau) = \langle b_\tau(t)^n \rangle \sim \tau^{\zeta(n)}$ exhibits a linear (monofractal) scaling, i.e., $\zeta(n) = nH$, with a single Hurst exponent $H$ [37]. It means that in contrast to large magnetohydrodynamic scales (with anomalous scaling), magnetic turbulence on small kinetic scales is not intermittent. Interestingly, several analyses based on Cluster data (we refer to the ESA's Cluster mission) [38,39] report *scale-invariant* behavior, while other analyses support strongly increasing scale-dependent kurtosis and associated departures from self-similarity [40,41] reports that within the kinetic scales, there is a general trend towards large kurtosis at smaller scales.

## V. CONCLUSIONS

The Magnetospheric Multiscale mission with unprecedented high millisecond time resolution of magnetometer data allows us to investigate turbulence on very small kinetic scales [19,20]. In these papers, we looked at the MMS observations above 20 Hz, where the magnetic spectrum becomes very steep, with the slope close to $-16/3$, possibly resulting from interaction between coherent structures [14].

Now, we have also taken into account plasma data with somewhat lower resolution for ion velocity. Following our previous studies in the inertial region [10,11], we have primarily shown that the Chapman-Kolmogorov equation, which is a necessary condition for the Markovian character of fluctuations, is satisfied for the transfer of energy between the magnetic and velocity fields also on much smaller kinetic scales. In fact, we have proved that a local transfer mechanism is present for transfer of ion velocity-to-velocity, velocity-to-magnetic, and magnetic-to-velocity stochastic fluctuations. This physically means that this energy transfer to smaller scales has a *local* character.

Moreover, we have verified that in the case of magnetic fluctuations, Kramers-Moyal expansion stops after the second term, resulting in the Fokker-Planck equation, with *drift* and *diffusion* terms, at least for scales smaller than (a) $\tau \sim 0.8$ s near the bow shock (BS), (b) $\tau \sim 9$ s inside the magnetosheath (SH), and (c) $\tau \sim 0.8$ s near the magnetopause (MP), correspondingly.

Similarly as for Parker Solar Probe (PSP) data [42], the lowest-order coefficients are linear and quadratic functions of magnetic fluctuations, which correspond to the generalized Ornstein-Uhlenbeck processes. As expected for some moderate scales, we have the Kappa distributions with heavy tails. The observed global universal scale invariance corresponds to a simple *linear* (monofractal) scaling. The extracted values of the nonextensity parameter in Tsallis entropy $q < 1$ equal to 0.341 in case (a), 0.273 in case (b), and (c) 0.716, respectively, provide robust measures of the departure of the system from equilibrium.

On the other hand, for velocity fluctuations, higher-order moments should be taken into account with possibly more complex dependence on the increments of velocity. Therefore, in this case, we also see somewhat more complicated probability density functions. Hence, unlike to the magnetic fluctuations, monoscaling does not occur, suggesting rather *nonlinear* (multifractal) scaling.

Nevertheless, we are still hoping that our observation of Markovian features in solar wind turbulence will be important for understanding the relationship between deterministic and stochastic properties of turbulence cascade on kinetic scales in complex physical systems.

## ACKNOWLEDGMENTS

We thank Marek Strumik for discussion on the theory of Markov processes. We are grateful for the efforts of the entire MMS mission, with J. L. Burch, Principle Investigator, C. T. Russell and the magnetometer team, including development, science operations, and the Science Data Center at the University of Colorado. The magnetic field data from the magnetometer are available online [22]. We acknowledge B. L. Giles, Project Scientist, for information about the magnetic field instrument, and D. G. Sibeck and M. V. D. Silveira for discussions during previous visits by W.M.M. to the NASA Goddard Space Flight Center. The data have been processed using the statistical programming language R. We would like to thank the referee for useful comments in relation to the problem of scale invariance of magnetic turbulence. This work has been supported by the National Science Centre, Poland (NCN), through Grant No. 2021/41/B/ST10/00823.

## APPENDIX A: TIME SERIES CHARACTERISTICS

The analysis of time series is significant in nonlinear dynamics because a thorough investigation can reveal all the essential information on the dynamical properties of the considered system. To estimate the target variable in predicting or forecasting, one uses the time variable (or in our case, also a timescale) as the reference point. The *time series analysis* involves the examination of the characteristics of the selected variables as a function of time, considered as the independent variable. Time series data for our investigation can be directly extracted from the measurements within the MMS mission.

A basic assumption regarding the data type is referred to as stationarity. Precisely, the time series is said to be stationary if it exhibits no trend, no seasonality, has a constant variance over time, and a consistent autocorrelation function over time. To evaluate this feature, one can plot the data and visually look for trend and seasonal components, although the more robust methods include statistical tests, such as the Augmented Dickey-Fuller (ADF) test [29], and the Kwiatkowski-Phillips-Schmidt-Shin (KPSS) test [30]. The ADF test uses the following hypotheses:

(i) $H_0$: the time series is nonstationary vs

(ii) $H_1$: the time series is stationary.

Note that it is set up this way to maintain a skeptical and cautious approach towards the findings. The null hypothesis is assumed to be true until the data present sufficient evidence that it is not. Naturally, when the $p$ value is less than a prechosen significance level, say $\alpha = 0.05$ or, more strictly, 0.01, then the null hypothesis can be rejected, which leads to the conclusion that the time series is stationary. The test for nonstationarity of a time series $\{y_t\}$ observed over $T$ time periods is estimated in the ADF regression model,

$$\Delta y_t = \alpha + \beta t + \gamma y_{t-1} + \sum_{j=1}^{p} \rho_j \Delta y_{t-j} + \varepsilon_t, \quad \text{(A1)}$$

where $\alpha$ is a constant, $\beta$ is a coefficient on a time trend, $p$ is a lag order of the *autoregressive* (AR) process, $\varepsilon_t$ is an error term (assumed white noise), and differencing terms $\Delta y_{t-j} = y_{t-j} - y_{t-j-1}$, with coefficients $\rho_j$. Here in the representation of ADF, the differencing term is added, in contrast to the standard Dickey-Fuller test. Once we get a value for the test statistic $\text{DF}_t = \hat{\gamma}/\text{SE}(\hat{\gamma})$, where SE denotes the standard error of the estimator, it can be compared to the relevant critical value for the Dickey-Fuller test.

Alternatively, the statistical Kwiatkowski-Phillips-Schmidt-Shin (KPSS) test is a type of unit root test. A unit root is a stochastic trend in a time series that can cause problems in statistical inference. The KPSS test is used for assessing the stationarity of a series around a deterministic trend. Namely, we have the following hypotheses:

(i) $H_0$: the time series is trend stationary or has no unit root vs

(ii) $H_1$: the time series is nonstationary or has a unit root.

Both the ADF and KPSS tests are frequently used to examine the stationarity of times series. A significant distinction between the ADF test and the KPSS test lies in their null hypotheses. In the KPSS test, the null hypothesis assumes stationarity for the series, while for the ADF test, it implies nonstationarity. Consequently, the practical interpretation of the *p* value differs in a contrasting manner for these two tests. A *p* value below a given significance level ($\alpha = 0.05$ or $0.01$) implies nonstationarity, whereas for the ADF test, such a *p* value indicates stationarity of the tested series.

Nevertheless, nonstationary time series can be converted into stationary, e.g., using a differencing method, which is a simple transformation of the series. This process effectively mitigates the series dependence on time and stabilizes its mean, resulting in a reduction of trend and seasonality within the series. The simplest difference between successive observations is called a lag-1 difference. However, one should be careful not to over-difference the time series, as it may lose some important information or features.

It is also worth mentioning that the time series have some limitations. For instance, if the missing values are not supported, some specific tools are needed to effectively address this problem. To fill in the empty data spots, a linear interpolation method can be used. It is a process of estimating the value of a function at a specific point, based on the known values at neighboring points. Hence, it requires knowledge of two points and the constant rate of change between them. The primary distinction between an interpolation and regression methods lies in the requirement to precisely match all data points in interpolation, whereas regression does not demand such an exact fit.

## APPENDIX B: DATA TRANSFORMATIONS

To compare the cPDF and use the CK condition (4), some preliminary data transformations are necessary. As is known, the *feature engineering* holds a vital role in the development of data analysis. These features can be time, categorical, and continuous variables. Among the various available techniques, we highlight the *feature binning* method. It is used for the transformation of a continuous or numerical variable into a categorical feature. Binning continuous variables may introduce nonlinearity and typically enhances the model performance. We have employed here a specific binning approach known as *equal-width unsupervised binning*. This method falls within the category of binning techniques that converts numerical or continuous variables into categorical bins, without taking the target class label into consideration. This algorithm segments the continuous variable into multiple categories, each characterized by bins or ranges of equal width. To be precise, in our case, we have begun by estimating the empirical joint PDFs, denoted as $P(x_1, \tau_1; x_2, \tau_2)$. This estimation process involved counting the occurrences of distinct pairs $(x_1, x_2)$ on a two-dimensional grid with equally spaced data bins, each of moderate size. Next we have carried out the normalization to ensure that the integral over all bins is equal to one. In a similar way, we have estimated the empirical one-dimensional PDFs $P(x_2, \tau_2)$ using a one-dimensional grid of bins, followed by the normalization. Finally, to obtain the empirical cPDFs, we have precisely applied the conditional probability formula numerically. The advantage of this approach is that it is simple to implement and interpret, and it preserves the distribution of the analyzed data.

In the same way, to distinguish between the local and nonlocal behavior of fluctuations between two quantities *x* and *y* (instead one of these), one needs to estimate the empirical cPDFs by using a generalized CK condition (5). Therefore, because of the difference in time resolutions of both analyzed time series, we must address the grid of time dimension. We can employ either the method of mean downsampling (decimation) or upsampling (interpolation) on the chosen variable. Contrary to a simple linear interpolation briefly described in Appendix A, decimation aims to reduce the original sample rate of the input signal to a lower rate by an integer factor. This factor is just a ratio of the input rate to the output rate. To increase the clarity of the presented graphs and ensure better interpretation of the results, we have opted for double-logarithmic (log-log) and semilogarithmic (log-linear) scales in some specific instances, rather than the standard linear scales. This adjustment of a double-logarithmic scale allows power-law representations to come out as straight lines in the spectra. Additionally, the semilogarithmic scale has been employed in Figs. 4–6, and 9.

- [1] U. Frisch, *Turbulence. The Legacy of A. N. Kolmogorov* (Cambridge University Press, Cambridge, 1995).
- [2] D. Biskamp, *Magnetohydrodynamic Turbulence* (Cambridge University Press, Cambridge, 2003).
- [3] R. Bruno and V. Carbone, *Turbulence in the Solar Wind*, Lecture Notes in Physics (Springer International, Berlin, 2016), Vol. 928.
- [4] L. F. Burlaga, *Interplanetary Magnetohydrodynamics*, International Series on Astronomy and Astrophysics (Oxford University Press, New York, 1995).
- [5] R. A. Treumann, Astron. Astrophys. Rev. **17**, 409 (2009).
- [6] A. Kolmogorov, Akademiia Nauk SSSR Doklady **30**, 301 (1941).
- [7] R. H. Kraichnan, Phys. Fluids **8**, 1385 (1965).
- [8] C. Renner, J. Peinke, and R. Friedrich, J. Fluid Mech. **433**, 383 (2001).
- [9] G. Pedrizzetti and E. A. Novikov, J. Fluid Mech. **280**, 69 (1994).
- [10] M. Strumik and W. M. Macek, Phys. Rev. E **78**, 026414 (2008).
- [11] M. Strumik and W. M. Macek, Nonlinear Process. Geophys. **15**, 607 (2008b).
- [12] R. Friedrich, J. Peinke, M. Sahimi, and M. Reza Rahimi Tabar, Phys. Rep. **506**, 87 (2011).
- [13] P. Rinn, P. Lind, M. Wächter, and J. Peinke, J. Open Res. Software **4**, 34 (2016).
- [14] W. M. Macek, A. Krasińska, M. V. D. Silveira, D. G. Sibeck, A. Wawrzaszek, J. L. Burch, and C. T. Russell, Astrophys. J. Lett. **864**, L29 (2018).

[15] F. Sahraoui, M. L. Goldstein, P. Robert, and Y. V. Khotyaintsev, Phys. Rev. Lett. **102**, 231102 (2009).

[16] O. Stawicki, S. P. Gary, and H. Li, J. Geophys. Res. **106**, 8273 (2001).

[17] Y. Narita, Astrophys. J. **831**, 83 (2016).

[18] A. A. Schekochihin, S. C. Cowley, W. Dorland, G. W. Hammett, G. G. Howes, E. Quataert, and T. Tatsuno, Astrophys. J. Suppl. Ser. **182**, 310 (2009).

[19] W. M. Macek, D. Wójcik, and J. L. Burch, Astrophys. J. **943**, 152 (2023).

[20] W. M. Macek and D. Wójcik, Mon. Not. R. Astronom. Soc. **526**, 5779 (2023).

[21] J. L. Burch, T. E. Moore, R. B. Torbert, and B. L. Giles, Space Sci. Rev. **199**, 5 (2016).

[22] http://cdaweb.gsfc.nasa.gov.

[23] G. I. Taylor, Proc. R. Soc., London Ser. A **164**, 476 (1938).

[24] H. Risken, *The Fokker-Planck Equation: Methods of Solution and Applications*, Springer Series in Synergetics (Springer, Berlin, 1996).

[25] M. Tabar, *Analysis and Data-Based Reconstruction of Complex Nonlinear Dynamical Systems: Using the Methods of Stochastic Processes* (Springer, Cham, 2019), p. 280.

[26] F. B. Hanson, *Applied Stochastic Processes and Control for Jump-Diffusions: Modeling, Analysis and Computation* (Society of Industrial Applied Mathematics, Philadelphia, PA, 2007).

[27] M. Anvari, M. Tabar, Peinke, and K. J. Lehnertz, Sci. Rep. **6**, 35435 (2016).

[28] E. Daly and A. Porporato, Phys. Rev. E **73**, 026108 (2006).

[29] D. A. Dickey and W. A. Fuller, J. Am. Stat. Assoc. **74**, 427 (1979).

[30] D. Kwiatkowski, P. C. Phillips, P. Schmidt, and Y. Shin, J. Econometr. **54**, 159 (1992).

[31] S. Lück, C. Renner, J. Peinke, and R. Friedrich, Phys. Lett. A **359**, 335 (2006).

[32] S. Benella, M. Stumpo, T. Alberti, O. Pezzi, E. Papini, E. Yordanova, F. Valentini, and G. Consolini, Phys. Rev. Res. **5**, L042014 (2023).

[33] W. M. Macek, Physica D **122**, 254 (1998).

[34] W. M. Macek and S. Redaelli, Phys. Rev. E **62**, 6496 (2000).

[35] C. Tsallis, J. Stat. Phys. **52**, 479 (1988).

[36] L. F. Burlaga and A. F. Viñas, Physica A **356**, 375 (2005).

[37] H. E. Hurst, R. P. Black, and Y. M. Simaika, *Long-term Storage: An Experimental Study*, Vol. 129 (Constable, London, 1965), pp. 591–593.

[38] K. H. Kiyani, S. C. Chapman, Y. V. Khotyaintsev, M. W. Dunlop, and F. Sahraoui, Phys. Rev. Lett. **103**, 075006 (2009).

[39] K. H. Kiyani, S. C. Chapman, F. Sahraoui, B. Hnat, O. Fauvarque, and Y. V. Khotyaintsev, Astrophys. J. **763**, 10 (2013).

[40] O. Alexandrova, V. Carbone, P. Veltri, and L. Sorriso-Valvo, Astrophys. J. **674**, 1153 (2008).

[41] P. Wu, S. Perri, K. Osman, M. Wan, W. H. Matthaeus, M. A. Shay, M. L. Goldstein, H. Karimabadi, and S. Chapman, Astrophys. J. Lett. **763**, L30 (2013).

[42] S. Benella, M. Stumpo, G. Consolini, T. Alberti, V. Carbone, and M. Laurenza, Astrophys. J. Lett. **928**, L21 (2022).